\documentclass[twocolumn,aps,showpacs]{revtex4}
\usepackage{amsmath}
\usepackage{amssymb}
\usepackage{graphicx}
\usepackage{dcolumn}
\usepackage{graphicx}
\usepackage{dcolumn}
\usepackage{bm}

\begin{document}

%\preprint{Phys.Rev.B}

\title{Unconventional Hall effect near charge neutrality point
in a two-dimensional electron-hole system}
\author{O. E. Raichev,$^1$ G.M.Gusev,$^2$ E.B Olshanetsky,$^3$  Z.D.Kvon,$^3$
N.N.Mikhailov,$^3$ S.A.Dvoretsky,$^{3}$ and J. C. Portal$^{4,5,6}$}

\affiliation{$^1$Institute of Semiconductor Physics, NAS of Ukraine,
Prospekt Nauki 41, 03028, Kiev, Ukraine} \affiliation{$^2$Instituto
de F\'{\i}sica da Universidade de S\~ao Paulo, 135960-170, S\~ao
Paulo, SP, Brazil} \affiliation{$^3$Institute of Semiconductor
Physics, Novosibirsk 630090, Russia} \affiliation{$^4$LNCMI-CNRS,
UPR 3228, BP 166, 38042 Grenoble Cedex 9, France}
\affiliation{$^5$INSA Toulouse, 31077 Toulouse Cedex 4, France}
\affiliation{$^6$Institut Universitaire de France, 75005 Paris,
France}

\date{\today}
\begin{abstract}
The transport properties of the two-dimensional system in HgTe-based quantum wells
containing simultaneously electrons and holes of low densities are examined. The Hall
resistance, as a function of perpendicular magnetic field, reveals an unconventional
behavior, different from the classical N-shaped dependence typical for bipolar systems
with electron-hole asymmetry. The quantum features of magnetotransport are explained
by means of numerical calculation of Landau level spectrum based on Kane Hamiltonian.
The origin of the quantum Hall plateau $\sigma_{xx}=0$ near the charge neutrality
point is attributed to special features of Landau quantization in our system.


\end{abstract}

\maketitle

\section{Introduction}

The renewed interest to the study of the integer quantum Hall effect (QHE)
has been manifested recently in investigation of the anomalous QHE state
in graphene, which provides electron or hole excitations near the Dirac
point of double-cone energy spectrum \cite{novoselov, zhang}. %\cite{1,2}.
It has been predicted and observed that quantized values of the Hall
conductance $\sigma_{xy}= \nu e^2/h$ correspond to filling factors
$\nu=4 (n+1/2)$, where $n=0, \pm 1, \pm 2, ...$ are integers
and the factor of 4 accounts for double degeneracy in both spin and valley
numbers. The half-integer form reflects a specific property of Landau
quantization for massless Dirac fermions. In particular, zero Landau level,
whose position coincides with the Dirac point, is composed from half hole
and half electron states, so the Hall conductance indicates a smooth
transition from hole-like (positive $\nu$) to electron-like (negative
$\nu$) behavior as the Fermi energy goes up through this point. In very
strong magnetic fields the spin degeneracy (and, possibly, the valley
degeneracy) is lifted, which means opening the gap at the Dirac point.
Theoretical models considering electron transport in these conditions can be
classified in two groups: quantum Hall metal (spin-first lifting scenario)
and quantum Hall insulator (valley-first lifting scenario) \cite{sarma}.
In the spin-first scenario, there exists a pair of counterpropagating chiral
edge states (with opposite spins) in the gap so that a quantized Hall state
at $\nu=0$ appears \cite{abanin, wiedmann, dean}. These states provide
a dominant contribution to the conductivity, while the bulk transport is suppressed
by the energy gap. This leads to divergence of the longitudinal resistivity $\rho_{xx}$
and smooth zero crossing of the Hall resistivity $\rho_{yz}$. In the valley-first
scenario, no edge states exist in the gap and the divergence of both $\rho_{xx}$
and $\rho_{yx}$ at $\nu=0$ is expected.

Since an unconventional quantum Hall state at $\nu=0$ does not rely on relativistic
dispersion of excitation, which is a specific case of graphene, it can be realized
in other materials where two-dimensional (2D) electrons and holes coexist. The wide
CdHgTe/HgTe/CdHgTe quantum wells, where separation of the size-quantized subbands
is relatively small, are of particular interest in this connection \cite{kvon}.
The 2D conduction (c) band in such systems is formed from the first heavy-hole subband
($h1$) whose effective mass is positive \cite{dyakon}, while the second heavy-hole subband
($h2$) form the upper part of the 2D valence (v) band. Owing to a uniaxial strain
of HgTe layer, which is caused by the lattice mismatch of HgTe and CdHgTe \cite{brune},
the energy spectrum of the $h2$ subband is essentially non-monotonic and has maxima
away from the $\Gamma$ point of the 2D Brillouin zone \cite{kvon112}. Depending
on the well width, strain strength, and interface orientation, the band structure
can be of the two following kinds: indirect-gap 2D semiconductor and 2D semimetal (Fig. 1),
which differ, respectively, by the absence or presence of the overlap of $h1$ and $h2$
subbands. In both these cases, a variation of the Fermi energy may cause a transition
between hole-like ($\sigma_{xy}>0$) and electron-like ($\sigma_{xy}<0$) behavior,
as is understandable even from a classical transport picture. Indeed, multiple intersections
of the Fermi level with non-monotonic energy spectrum (Fig. 1) lead to a complicated
Fermi surface for 2D electrons (Fermi arc) composed of more than one closed branches
providing both electron-like and hole-like orbits in the presence of the perpendicular
magnetic field \cite{kittel}.

\begin{figure}[ht]
\includegraphics[width=7cm,clip=]{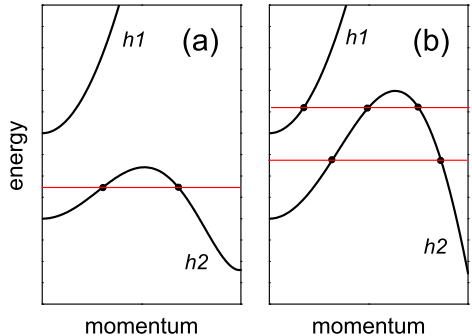}
\caption{\label{fig.1} (Color online) A schematic representation of two kinds
of band structure in a wide HgTe quantum well: indirect-gap 2D semiconductor (a)
and 2D semimetal (b). The straight lines show the position of the Fermi level for
the cases when the Fermi arc contains more than one branch.}
\end{figure}

Recently, it has been demonstrated that the Hall conductivity of CdHgTe/HgTe/CdHgTe
quantum wells of 20 nm width and [013] interface direction exhibits a quantized plateau
$\sigma_{xy}=0$ in the magnetic fields of a few Tesla when the gate voltage is varied
in the vicinity of the charge neutrality point (CNP) \cite{gusev}. The universality
of the edge-state transport picture suggests the existence of a pair of counterpropagating
edge states in such 2D systems, which is equivalent to QHE near the Dirac point
in graphene within a spin-first scenario. Despite of this similarity, one should point
out several differences. Unlike the case of graphene, the observation of the QHE state with
$\sigma_{xy}=0$ in HgTe quantum wells does not require ultrahigh magnetic fields. More
important, the band structure of HgTe quantum wells is very different from that of graphene.
Graphene is a gapless 2D material with symmetric and monotonic energy spectrum of electrons
and holes. The $n = 0$ Landau level resides precisely at the electron-hole symmetric point
(Dirac point), which allows for a $\sigma_{xy}=0$ QHE when the the 4-fold (spin and valley)
degeneracy is lifted. In wide HgTe quantum wells the energy spectrum of v-band is
non-monotonic, and the extrema of the c-band and v-band are shifted in momentum space with
respect to each other. The simplest consequence of such asymmetry is that zero crossing of
the Hall resistivity takes place away from the CNP, as shown below. The overlap of c-band
and v-band makes the transport picture even more complicated. The question about mechanisms
leading to opening of the gap responsible for the $\sigma_{xy}=0$ QHE in wide HgTe
quantum wells requires further condideration.

In this paper, we present experimental results on the Hall resistivity in 20 nm wide
HgTe-based quantum wells with [001] interface orientation containing simultaneously
electrons and holes. The densities of carriers in these wells are considerably
smaller those those for the [013]-, [112]- and slightly wider [001]-grown wells examined
previously in our experiments \cite{kvon}, \cite{kvon112}, \cite{gusev}, \cite{olshanetsky}.
For this reason, we see quantum features in transport at smaller magnetic fields. Apart from
the existence of the $\sigma_{xy}=0$ plato in the dependence of Hall conductivity on the gate
voltage, we have found an unusual non-monotonic dependence of the Hall resistivity $\rho_{yx}$
on the magnetic field. This dependence essentially differs from the classical N-shaped
Hall resistivity expected for electron-hole systems. A theoretical consideration based on
calculation of the Landau level spectrum for our structure qualitatively explains the main
features of our observations and helps us to uncover a mechanism of transition to the
$\sigma_{xy}=0$ state in the systems under investigation.

Tha paper is organized as follows. In section II we describe the details of the experiment
and give experimental results. Section III presents a theoretical basis for consideration
of magnetotransport in 2D semimetals, Landau level calculation for wide HgTe quantum wells,
and discussion of the results. The conslusions are briefly stated in the final section.

\section{Experiment}

The Cd$_{0.65}$Hg$_{0.35}$Te/HgTe/Cd$_{0.65}$Hg$_{0.35}$Te quantum wells with [001] surface
orientations and the width of 20 nm were prepared by molecular beam epitaxy. A detailed
description of the sample structure has been given in Refs. \cite{kvon, gusev, olshanetsky}.
The top view of a typical experimental sample is shown in the inset to Fig. 2 (a). The
sample consists of three 50 $\mu$m wide consecutive segments of different length (100,
250, and 100 $\mu$m), and 8 voltage probes. The ohmic contacts to the 2D layer were formed
by the in-burning of indium. To prepare the gate, a dielectric layer containing 100 nm SiO$_{2}$
and 200 nm Si$_{3}$Ni$_{4}$ was first grown on the structure using the plasmochemical
method. Then, the TiAu gate was deposited. The rate of the density variation with gate
voltage is estimated as $\alpha=1.09 \times 10^{15}$ m$^{-2}$ V$^{-1}$. The magnetotransport
measurements in these structures were performed in the temperature range 0.8-10 K and
in magnetic fields up to 5 T using a standard four-point circuit with a 3-13 Hz alternating
current of 1-10 nA through the sample, which is sufficiently low to avoid the overheating
effects. Several devices from the same wafer have been studied. \\ \\ \\

\begin{figure}[ht]
\includegraphics[width=8cm,clip=]{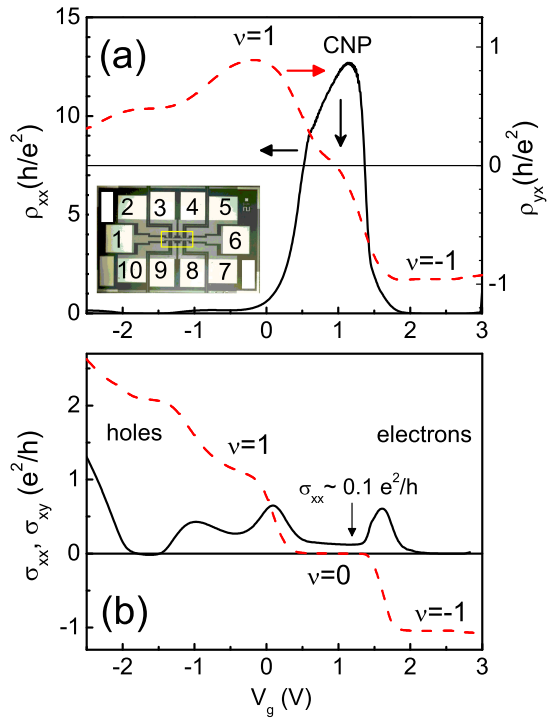}
\caption{\label{fig.2}(Color online) (a) The longitudinal $\rho_{xx}$ and Hall $\rho_{yx}$
resistivities as a function of the gate voltage, T=0.9 K, B=5 T. Insert-top view of the sample.
The perimeter of the gate is shown by rectangle. (b) The longitudinal $\sigma_{xx}$ and
Hall $\sigma_{xy}$ conductivities as a function of the gate voltage, $T=0.9$ K, $B=3.75$ T.
The arrow indicates the position of the charge neutrality point, where $n_{e}= n_{h}$.}
\end{figure}

The longitudinal resistivity $\rho_{xx}$ and corresponding Hall resistivity $\rho_{yx}$ acquired
by varying the gate voltage at a constant magnetic field $B=5$ T are shown in Fig. 2 (a). Figure 2 (b)
shows the the longitudinal conductivity $\sigma_{xx}$ and Hall conductivity $\sigma_{xy}$ calculated
from the experimentally measured resistivities by tensor inversion. Possible admixtures of longitudinal
and Hall resistivities, caused by contact misalignment and inhomogeneities, were removed, by symmetrizing
all measured values for positive and negative magnetic fields. We may see that calculating the conductivities
from the magnetoresistance peak at the CNP and a zero crossing of the Hall resistance yields a zero minimum
in the longitudinal conductivity $\sigma_{xx}$ and a quantized zero plateau in the Hall conductivity
$\sigma_{xy}$ at $\nu=0$. This behaviour, observed in the fields above 3.6 T, agrees with our previous
study of the quantum Hall effect near CNP in wide HgTe quantum wells in samples with [013] surface
orientation \cite{gusev}.

\begin{figure}[ht]
\includegraphics[width=8cm,clip=]{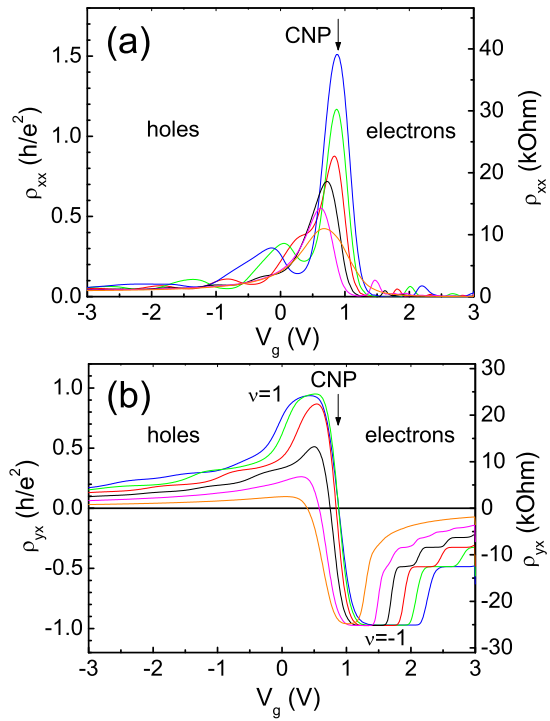}
\caption{\label{fig.3} (Color online) The longitudinal $\rho_{xx}$
(a) and Hall $\rho_{yx}$ (b) resistivity at T=0.9 K as a function of
the gate voltage for different magnetic fields $B$ (T): 0.5 (orange),
1 (magenta), 1.5 (black), 2.0 (red), 2.5 (green), 3 (blue), .}
\end{figure}

Now we turn our attention to magnetoresistance measured as a
function of the gate voltage with increasing magnetic field, shown
in Fig. 3. For higher voltages, corresponding to electron-like
conductivity, all the quantized plateaux are already developed in
the field of 1 T. In the field of 0.5 T, however, we see only a
short plateau at $\nu=-1$ and a weak indication of $\nu=-3$ plateau.
For lower voltages, corresponding to hole-like conductivity, we do
not see any well-developed plateaux up to 3 T, although above 1.5 T
there is periodic flattening of the Hall resistance picture possibly
suggesting that the effects of Landau quantization become also
important for holes. Surprisingly, the peak of the resistivity and
the smooth zero crossing point of the Hall resistivity, which has
been identified previously with CNP, both are shifted to higher
positive voltage with increasing $B$. At weak magnetic fields this
shift is linear in $B$, while at higher fields a saturation is
reached (no more shift occurs).

\begin{figure}[ht]
\includegraphics[width=9cm,clip=]{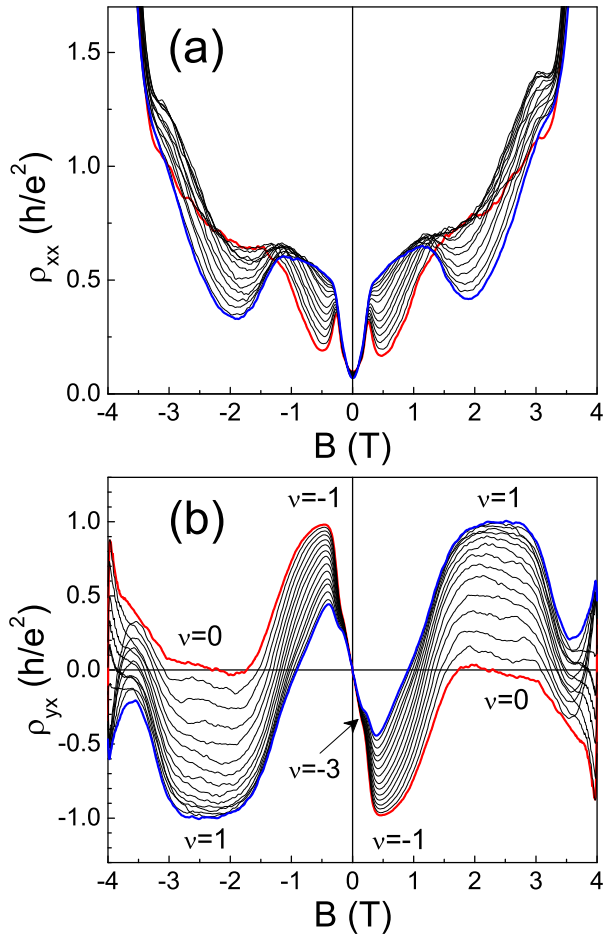}
\caption{\label{fig.4} (Color online) The longitudinal $\rho_{xx}$ (a)
and Hall $\rho_{yx}$ (b) resistivities as functions of the magnetic field
for different gate voltages near CNP, $T=0.9$ K. The highest and the lowest
gate voltages are shown by broader (colored) lines. The red line corresponds
to CNP ($n_e \simeq n_h$), while the blue line corresponds to $n_h-n_e
\simeq 1.6 \times 10^{10}$ cm$^{-2}$.}
\end{figure}

The most intriguing behavior is observed in the magnetic-field dependence of the Hall resistivity.
In Fig. 4 (a,b) we give detailed plots of $\rho_{xx}$ and $\rho_{yx}$ for different gate voltages
near CNP. Instead of a smooth N-shaped dependence of the Hall resistivity suggested by a classical
magnetotransport theory for semimetals (see Sec. III) and observed in our previous experiments, we
see a non-monotonic dependence showing a number of remarkable features discussed
below. Up to the fields of $0.4$ T, the Hall resistivity behaves nearly linear with $B$, demonstrating,
however, a weak shoulder identified as a quantized plateau $\rho_{yx}=-h/3e^2$. At $B \simeq 0.45$ T
we see a short plateau where the Hall resistivity almost reaches the resistance quantum [$\rho_{yx}
\simeq 0.92(-h/e^2)$], which is accompanied by a deep minimum in $\rho_{xx}$ in agreement with
conventional quantum Hall behaviour in unipolar ($n$-type) system. With further increase in magnetic
field ($B > 0.5$ T) the absolute value of the Hall resistivity drops down. In the interval 2 T $< B
<$ 3 T there appear new features resembling the plateaux with $\rho_{yx}$ ranging from 0 to $h/e^2$
in the chosen interval of gate voltages. The plateau $\rho_{yx}=h/e^2$ corresponds to conventional
quantum Hall effect in hole ($p$-type) system, and is accompanied by a minimum in $\rho_{xx}$. However,
these Hall plateaux do not exhibit exact quantization, as is typical for the states not yet fully
formed in the magnetic field. Moreover, the value of $\rho_{xy}$ smoothly changes with the gate
voltage, which possibly indicates that the Fermi level does not lie in the region of localized hole
states between Landau levels and the contribution of bulk delocalized states to transport is essential.
Finally, in the region above 3.5 T $\rho_{xy}$ demonstrates a complicated non-monotonic behavior
which strongly depends on the gate voltage, while the resistivity $\rho_{xx}$ starts to grow
up sharply.

The observed features of Hall resistance and logitudinal resistivity are discussed in the
next section using both classical and quantum approaches to magnetotransport in 2D semimetals.

\section{Theory and discussion}

The calculated energy spectrum of carriers in the quantum well investigated in our experiment
is shown in Fig. 5. The calculations are based on a numerical solution of the eigenstate problem
for $6 \times 6$ matrix Kane Hamiltonian under approximation of symmetric rectangular confining
potential of the Cd$_{0.65}$Hg$_{0.35}$Te/HgTe/Cd$_{0.65}$Hg$_{0.35}$Te heterostructure. We used
the following parameters. The conduction- and valence-band offsets of the heterostructure under
investigation are $U_c=0.84$ eV and $U_v=0.36$ eV, respectively, the gap energy in HgTe is
$\varepsilon_g=-0.3$ eV, the spin-orbit splitting energy (both in HgTe and Cd$_{0.65}$Hg$_{0.35}$Te)
is $\Delta=1.0$ eV. The Kane matrix element $P$ and the Luttinger parameters ($\gamma_1$, $\gamma_2$,
and $\gamma_3$) of HgTe and CdTe are listed in the table of Ref. \cite{novik}. The main effect of
the uniaxial strain of HgTe layer was taken into account by introducing a shift $E_{st}$ of the
light-hole energy with respect to the heavy-hole energy in the Hamiltonian \cite{brune}. We use
$E_{st}=15$ meV, based on the quantity of $E_{st}=22$ meV given in Ref. \cite{brune} for HgTe on
pure CdTe substrate. The calculations demonstrate that our system is a 2D semimetal. Owing to a
difference in the Luttinger parameters $\gamma_2$ and $\gamma_3$, the spectrum of
the 2D v-band is anisotropic. This leads to formation of four v-band valleys along the
directions $[11]$, $[1\bar{1}]$, $[\bar{1}1]$, and $[\bar{1}\bar{1}]$ in the plane of 2D
momentum ${\bf p}=(p_x,p_y)$, with exterma at $p/\hbar \simeq 0.26$ nm$^{-1}$. The single
c-band valley with minimum at $p=0$ is almost perfectly isotropic. The corresponding five-branch
Fermi arc is shown in the inset. With lowering Fermi level $\varepsilon_F$, the picture of the
Fermi arc is modified. As $\varepsilon_F$ passes through the saddle points and, further,
through the c-band minimum at $p=0$, the Fermi arc becomes composed of, respectively, three
and two branches encircling the point $p=0$.

\begin{figure}[ht]
\includegraphics[width=9cm]{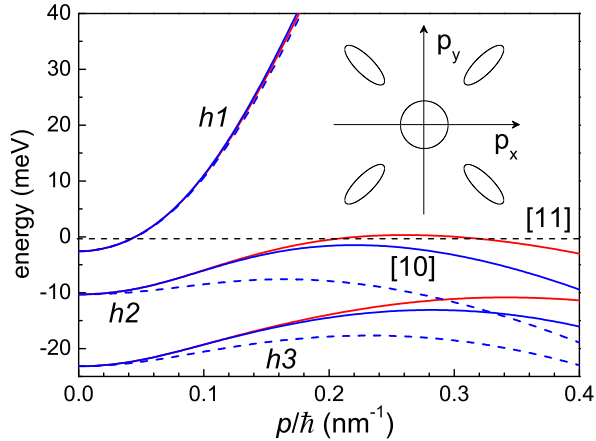}
\caption{\label{fig.5} (Color online) Calculated electron energy spectrum of a
20 nm symmetric Cd$_{0.65}$Hg$_{0.35}$Te/HgTe/Cd$_{0.65}$Hg$_{0.35}$Te quantum well
grown along [001] direction, for two direction of electron momenta in the 2D plane:
along [01] and [11]. Three upper heavy-hole ($h$) subbands are demonstrated. The dashed
curves (for [01] direction only) correspond to the calculation without strain.
The inset gives a schematic picture of the Fermi arc corresponding to the
Fermi level shown by a straight dashed line passing close enough to the valence-band
extrema. The area inside the circle and the total area inside the four ellipses
are proportional to electron and hole densities, $n_e$ and $n_h$, respectively.}
\end{figure}

Some features observed in our experiment, as well as in the previous experiments on 2D
semimetals (see refs. \cite{kvon}, \cite{kvon112}, \cite{olshanetsky}) can be understood
from consideration of the classical magnetoresistance of the system with energy spectrum
$\varepsilon_{\bf p}$ of semimetallic type shown in Fig. 5. The expressions for electron
and hole densities can be written as $n_e=2 S_e/(2 \pi \hbar)^2$ and $n_h=2 S_h/(2 \pi \hbar)^2$,
where the factor of 2 accounts for spin degeneracy and the quantities $S_i$ ($i=e,h$) are
given by the integrals over energy: $S_i=\int d \varepsilon s_i(\varepsilon)(-\partial
f_{\varepsilon}/\partial \varepsilon)$. Here $f_{\varepsilon}$ is the Fermi-Dirac
distribution function and $s_i(\varepsilon)$ is the area in the $(p_x,p_y)$ plane
encircled by the isoenergetic lines $\varepsilon=\varepsilon_{\bf p}$ under condition
that the sign of the transverse component of the group velocity for these lines is
either positive ($i=e$) or negative ($i=h$). At zero temperature, when
$\partial f_{\varepsilon}/\partial \varepsilon=-\delta(\varepsilon-\varepsilon_F)$,
$S_e$ and $S_h$ are the areas encircled by the Fermi arcs with positive and
negative transverse group velocities, respectively. Using this representation of $n_e$
and $n_h$, it can be demonstrated that the linear Hall conductivity in the limit of
classically strong magnetic field $B$, when scattering of the carriers during their
cyclotron motion is neglected, is written as \cite{kittel}
\begin{equation}
\sigma_{xy}=-|e| (n_e-n_h)/B.
%1
\end{equation}
This expression is valid for arbitrary energy spectrum and suggests that the Hall conductivity
changes its sign at CNP ($n_e=n_h$). However, at weaker $B$, when scattering is essential,
the point of zero $\sigma_{xy}$ is shifted away from the CNP. Indeed, the Hall conductivity
becomes sensitive not only to the signs but also to the absolute values of the group velocities
along the branches of the Fermi arc, because these velocities determine the densities of states
and scattering probabilities.

For arbitrary $B$, there is no exact solution of the kinetic equation describing the
magnetotransport in semimetals with anisotropic energy spectra. However, both
$\sigma_{xy}$ and $\sigma_{xx}$ can be written explicitly if the Fermi energy
is close to the valence-band extrema so the energy spectra everywhere in the vicinity
of $\varepsilon_{\bf p}=\varepsilon_F$ are approximated by parabolic functions.
In this case, which corresponds to the picture given in the inset of Fig. 5,
the components of the conductivity tensor at low temperature are the sums of
contributions from two groups of carriers near the exterma of the conduction
and valence bands (i.e., the electron and hole contributions):
\begin{equation}
\sigma_{xx}=\sigma^{(e)}_{xx} + \sigma^{(h)}_{xx},~~~\sigma_{xy}=\sigma^{(e)}_{xy} + \sigma^{(h)}_{xy},
%2
\end{equation}
\begin{equation}
\sigma^{(i)}_{xx} = \frac{|e| n_i \beta_i}{B^2+\beta^2_i},~~ \sigma^{(i)}_{xy}= \mp
\frac{|e| n_i B}{B^2+\beta^2_i},
%3
\end{equation}
where $\beta_i=m_i/|e| \tau_i$ are the inverse mobilities of electrons and holes expressed
through the corresponding effective masses $m_i$ and scattering times $\tau_i$. The
signs $-$ and $+$ in $\sigma^{(i)}_{xy}$ stand for electrons and holes, respectively. The
expresssions (2) and (3) turn out to be analogous to known expressions for two-component
electron-hole systems (see for example, Ref. \cite{seeger}). The introduction of the effective
mass $m_e$ is natural, since the c-band spectrum is isotropic and parabolic in a wide
energy range. The effective mass $m_h$ is introduced as $m_h=\sqrt{m_1 m_2}$, where
$m_1$ and $m_2$ are the masses characterizing the v-band energy spectrum along the main
axes of the ellipses. From our calculations (Fig. 5), we find $m_e =0.03$ $m_0$, $m_h \simeq 5$ $m_e$,
so the density of hole states (taking into account 4-fold valley degeneracy) is about 20 times
greaater than the density of electron states. This means that at low hole densities the
Fermi energy is pinned near the valence band extremum, and the parabolic approximarion for
$\varepsilon_{\bf p}$ in the valence band is justified. The expressions (2) and (3) lead to
the Hall resistivity in the form
\begin{eqnarray}
\rho_{yx}= -\frac{B}{|e|} \frac{(n_e-n_h)B^2 +(n_e \beta_h^2-n_h \beta_e^2)}{(n_e-n_h)^2 B^2
+ (n_e \beta_h+n_h \beta_e)^2}.
%4
\end{eqnarray}

Since $m_h \gg m_e$ and, consequently, $\beta_h \gg \beta_e$, the condition for zero $\rho_{yx}$
at low $B$ is realized at $n_h > n_e$, away from the CNP. The gate voltage $V_g^{(0)}$ corresponding
to zero crossing can be found from Eq. (4) combined with the expression $n_e-n_h = \alpha (V_g -
V_{g}^{CNP})$, where $\alpha$ is a constant specified for our sample in the previous section.
With increasing $B$, the voltage $V_g^{(0)}$ moves towards the CNP gate voltage $V_{g}^{CNP}$,
first linear with $B$, then saturating in the close vicinity of the CNP. Thus, the behavior of
the zero crossing point shown in Fig. 3 (b) follows from the classical expression (4).
The same expression describes N-shaped magnetic-field dependence of the Hall resistance at
fixed $V_g$, under condition that $n_h > n_e$. However, the Hall resistance plotted in Fig. 4 (b)
is different from the simple N-shaped dependence and cannot be described within the classical
theory.

To account for the quantum features of the magnetotransport in our system, it is important to
get an idea of how the energy spectrum is quantized in the magnetic field. To calculate the
Landau levels (LLs) of HgTe quantum wells, we have used isotropic approximation in the Kane
Hamiltonian, when the Luttinger parameters $\gamma_2$ and $\gamma_3$ both are
replaced by $(\gamma_2+\gamma_3)/2$. In this approximation, the problem is greatly simplified,
since columnar eigenstates of the $6 \times 6$ matrix Kane Hamiltonian become representable
in the form
\begin{equation}
\left( u^{1}_{n} \varphi_{n-2}, u^{2}_{n} \varphi_{n-1}, u^{3}_{n}
\varphi_{n-3}, u^{4}_{n} \varphi_{n-2}, u^{5}_{n} \varphi_{n-1},
u^{6}_{n} \varphi_{n} \right)^T
%5
\end{equation}
with the oscillatory functions $\varphi_{n}$ ($n \geq 0$) satisfying the relations
$\hat{k}_- \varphi_{n} = \varphi_{n-1}\sqrt{2n}/\ell$ and $\hat{k}_+ \varphi_{n} =
\varphi_{n+1}\sqrt{2(n+1)}/\ell$, $\ell=\sqrt{\hbar/|e|B}$ is the magnetic length.
The states are, therefore, described by the LL number $n$, and for each $n$ one can find
numerically the set of the components $u^{m}_{jn}(z)$ ($m=1,2,...6$) and the corresponding
discrete energies $\varepsilon_{jn}$, where the number $j$ accounts for both subband number
and spin state. These energies, for two chosen values of the magnetic field, are plotted
in Fig. 6. \cite{remark}. We point out the most essential properties of the LL spectrum.
There are two states for each subband with LL number $n \geq 3$, these states differ by projection
of spin. For $n=0$, 1, and 2 there is only one spin state within each heavy-hole subband, because the
third component of the wave function (5) (spin-up heavy-hole component) is missing for these
particular LL numbers. The levels originating from the c-band rapidly go up with increasing $B$,
except the first level (denoted as A), which slowly moves down with the rate approximately 0.25 meV/T.
In contrast, the levels originating from the v-band form a dense set, especially near the band extremum,
and are slowly shifted with increasing magnetic field. The v-band levels $n=1$ and $n=2$ (the latter
is denoted as B) are an exception, because they move up rather rapidly with increasing field.
At $B \simeq 2.8$ T the levels A and B cross. The existence of special LLs which belong to
different subbands and cross with increasing magnetic field is a common feature of the HgTe
quantum wells with inverted band spectrum, and such a behavior is confirmed experimentally
for narrower wells \cite{konig}, \cite{konigthes}. As the magnetic field increases further,
the level B becomes the upper one in the v-band, and a broad gap is formed at the v-band
exteremum, as seen in the right panel of Fig. 6.

\begin{figure}[ht]
%\centering
\includegraphics[width=9.5cm]{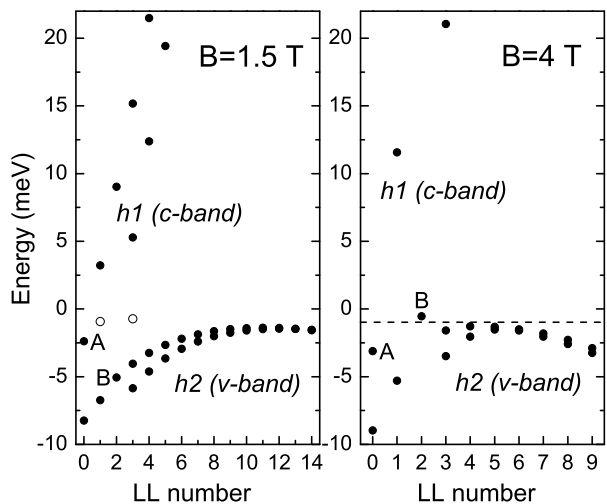}
\caption{\label{fig.6} Calculated Landau levels for 20 nm
symmetric Cd$_{0.65}$Hg$_{0.35}$Te/HgTe quantum well for $B=1.5$ T and $B=4$ T.
Only two sets of levels originating from two principal 2D subbands are shown.
The levels denoted as A (the first one in c-band) and B ($n=2$ level in v-band) experience
crossing at about 2.8 T. Starting from $B \simeq 3.6$ T, the level B becomes the upper
one in v-band. The empty circles in the left panel indicate position of the next two
c-band Landau levels at $B=0.3$ T. The dashed line in the right panel shows the
position of the Fermi energy corresponding to a quantized plateau $\sigma_{xy}=0$
near the CNP.}
\end{figure}

Based on the LL spectrum calculation, one may propose the following explanation of the
behavior of magnetoresistance shown in Fig. 4. The entire interval of the magnetic
fields investigated is subdivided in three regions: weak, intermediate, and strong
magnetic fields, see Fig. 7, which are discussed below.

In weak magnetic field region (up to 0.45 T), the $v$-band states are not quantized,
while the c-band states become quantized with increasing field because of the small effective
mass, $m_e = 0.03$ $m_0$, in this band. Since both electrons and holes are present in the
system, a fixed gate voltage means that $n_h-n_e$ is constant. At small differences $n_h-n_e$,
the Fermi level is pinned close to the v-band extremum, because of a large density of states in
the v-band. The QHE observed in this region of fields is caused by the passage of the $c$-band
LLs through the Fermi level. At $B \simeq 0.45$ T only the lowest c-band LL (level A) remains
occupied. Based on this scenario, we estimate the energy of semimetallic overlap as 1.1 meV.
Although this value is considerably smaller than the 3 meV overlap following from the numerical
calculations (Fig. 5), we do not find this discrepancy surprising, because the calculation
puts aside many unknown factors (such as deviation of the quantum well potential from a
simple rectangular one, partial relaxation of the strain inside the HgTe layer, etc.) which
may shift the quantization energies within several meV. In weaker fields, we observe a signature
of a plateau at c-band filling factor 3 ($\nu=-3$) but do not observe any plateau at $\nu=-2$.
This fact becomes understandable from our numerical calculation showing that two next to
level A c-band LLs are almost degenerate at low $B$, see empty circles in the left panel
of Fig. 6. Since these are the states with different spin numbers, the nature of the
degeneracy is described in terms of negligible spin splitting for low-lying c-band states
in weak magnetic fields.

The absence of any sizeable contribution of holes to transport at
CNP means that both longitudinal and Hall conductivities of holes are much
smaller that $e^2/h$. Though such small hole conductivities cannot be entirely described within
the classical (Drude) theory, it is likely that the holes experience strong localization.
With semimetallic overlap of 1.1 meV, the hole density at CNP is estimated as $n_h \simeq 1.2
\times 10^{10}$ cm$^{-2}$. Since the density of states in v-band is large, the presence of
disorder makes it possible that the majority of holes, at such small $n_h$, occupy the
tail of the density of states below the mobility threshold.

\begin{figure}[ht]
\includegraphics[width=8.5cm,clip=]{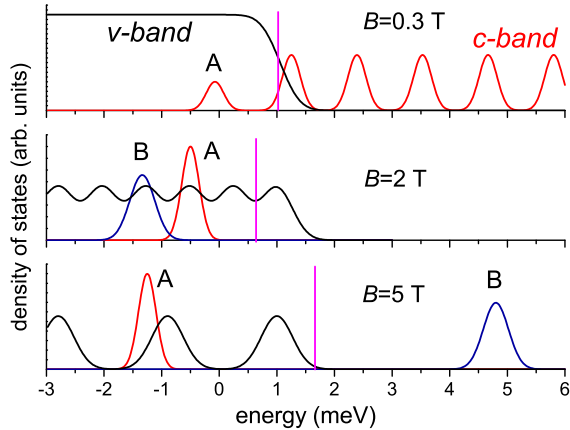}
\caption{\label{fig.7} (Color online) A schematic picture describing evolution
of the density of states and position of the Fermi level (straight vertical line)
in HgTe quantum wells with a small semimetallic overlap. Three regimes, corresponding
to weak, intermediate, and strong magnetic fields (from up to down), are emphasized
and discussed in the text. The c-band and v-band states are shown in different colors.
The v-band level B, which demonstrates a special behavior, is shown separately. The
energy is counted from the position of level A at $B \rightarrow 0$.}
\end{figure}

In the intermediate region (from 0.45 to 3 T), a single c-band level A
is occupied. The electron density $n_e$ increases linearly with the magnetic field.
Since the difference $n_h-n_e$ is fixed by the gate, the hole density $n_h$ increases as well,
and the holes become active in transport, causing a positive contribution to the Hall
conductivity, and, consequently, a positive deviation of the Hall resistivity from $-h/e^2$.
The increase of hole density by the gate leads to a similar effect. The components of the
conductivity tensor are described by Eq. (2) with constant electronic components $\sigma^{(e)}_{xx}
=0$ and $\sigma^{(e)}_{xy}=-e^2/h$ which can be treated also as contribution of the
chiral edge state originating from level A to transport. Thus, the dependence of
resistivity on magnetic field and gate voltage is determined by the hole components of
the conductivity tensor. The absence of a detailed knowledge of hole conductivity in the
intermediate magnetic field region does not allow us to describe the Hall resistance
quantitatively. Whereas the monotonic change of the Hall resistivity in the
region below 1.7 T might be explained within a classical representation of $\sigma^{(h)}_{xx}$
and $\sigma^{(h)}_{xy}$, the flat (plateau-like) features at higher fields possibly suggest
that Landau quantization of holes already starts and makes $\sigma^{(h)}_{xx}$ and
$\sigma^{(h)}_{xy}$ nearly independent on $B$ at $B > 1.7$ T. The fact that the value of
$\rho_{xy}$ still smoothly changes with the gate voltage everywhere in this region means
that the hole LLs are not well resolved, indeed these LLs form a dense set according to our
numerical calculations. The contribution of bulk delocalized hole states to transport is,
therefore, essential, and increases proportional to the hole density controlled by the gate.

In the high magnetic field region (above 3 T), the behavior of Hall and longitudinal resistances
is determined by another important factor, the influence of the v-band level B, which rapidly
goes up with increasing magnetic field. The reduction of the hole contribution to Hall resistivity
occurs when this level passes the Fermi level. According to our calculations, level B crosses
the upper one of the other v-band LLs approximately at $B=3.6$ T. As this happened, level B
itself becomes the upper one in the v-band and accumulates the majority of hole density. The
remaining fraction of hole density, equal to $n_h-n_e$, is much smaller than the capacity of
a single LL (at $B=4$ T the capacity of a single LL is about 10$^{11}$ cm$^{-2}$). Therefore,
the Fermi level resides in the region of localized states in the tail of the next v-band LL,
as shown in the lowest panel of Fig. 7, and bulk contribution to conductivity is negligible.
The transport is determined by a pair of counterpropagating chiral edge states originating
from LLs A and B, so the Hall conductivity $\sigma_{xy}$ is expected to be zero. Since small
variations of the gate voltage do not shift the Fermi level from the region of localized
states, a plateau $\sigma_{xy}=0$ appears in the gate-voltage dependence of $\sigma_{xy}$,
see Fig. 2 (b). The fact that we observe such a plateau starting from 3.6 T is
consistent with the calculated behavior of LLs and emphasizes the crucial role of the
level B in magnetotransport in wide (semimetallic) HgTe wells.

A description of the longitudinal and Hall resistivities in the regime of counterpropagating
edge states reqires a more detailed consideration \cite{abanin}, because in view of strong
scattering between these edge modes the edge-state transport is suppressed and the small
bulk contribution to conductivity might be important as well. We apply the
formalism developed in Ref. \cite{abanin} to a system comprising two edge
states of electron ($e$) and hole ($h$) kind and a continuum of bulk
states. The scattering between the edge states and the scattering between bulk states and
each of the edge states is characterized by mean free path lengths $\gamma^{-1}$, $g_e^{-1}$,
and $g_h^{-1}$, respectively, which are assumed to be smaller than the sample dimensions.
Then we find
\begin{equation}
\rho_{xx}=\left[\frac{1}{\rho^{(b)}_{xx}} + \frac{2e^2}{h L_y}
\frac{g_e+g_h}{\gamma(g_e+g_h)+g_e g_h} \right]^{-1},
%6
\end{equation}
where $L_y$ is the width of the sample, and $\rho^{(b)}_{xx}$ is the bulk resistivity expressed
by a standard tensor inversion through the bulk conductivities $\sigma^{(b)}_{xx}$ and
$\sigma^{(b)}_{xy}$. Since both these conductivities are much smaller than $e^2/h$ in the
regime under consideration, and backscattering is strong, $\gamma L_y \gg 1$, the
resistivity $\rho_{xx}$ becomes much larger than the resistance quantum, as seen
in Fig. 4 (a). The Hall resistance is
\begin{equation}
\rho_{yx}=\rho_{xx} \left[\frac{\sigma^{(b)}_{xy}}{\sigma^{(b)}_{xx}}+ \frac{1}{L_y}
\frac{g_h-g_e}{\gamma (g_e+g_h)+ g_e g_h} \right].
%7
\end{equation}
Depending on parameters, $\rho_{yx}$ may behave in different ways. Assuming for example,
that both $\sigma^{(b)}_{xx}$ and $\sigma^{(b)}_{xy}$ are exponentially small, one cannot say that
their ratio is small as well, so both contributions in the square brackets of Eq. (7) remain
essential. If, in addition, $\sigma^{(b)}_{xy}/\sigma^{(b)}_{xx} \ll (\gamma L_y)^{-1}$, one obtais
$\rho_{yx}=\mu h/2e^2$ with $\mu=(g_h-g_e)/(g_h+g_e)$. When $\sigma_{xy}$ is calculated from
the expressions (6) and (7) by tensor inversion, the large value of $\rho_{xx}$ guarantees
that $\sigma_{xy} \ll e^2/h$ regardless to behavior of $\rho_{yx}$.

\section{Conclusions}

The possibility to create gated HgTe wells of different widths opens the avenue for
investigation of bipolar 2D semimetals where electron and hole densities can be varied
in a wide range. In this paper, we have examined low-temperature magnetotransport in the
system with a very small semimetallic overlap, about 1 meV, which provides densities of
electrons and holes of the order $10^{10}$ cm$^{-2}$. We have demonstrated, both
experimentally and theoretically, that such a system has rather interesting properties
generally following from electron-hole asymmetry and special features of Landau quantization
in HgTe wells. The low density of carriers assures manifestation of quantum effects already
at $B < 0.5$ T. With increasing magnetic field $B$, there exist three distinct transport
regimes, where dependence of longitudinal and Hall resistances on $B$ is determined,
respectively, by electrons, holes, and a pair of counterpropagating edge states originating
from conduction and valence bands. Some of the features observed, in particular, a $B$-dependent
shift of the zero Hall resistance point away from the charge neutrality point, are explained
within the classical transport picture. The quantum features of magnetotransport are in agreement
with the calculated Landau level spectrum of our system. This calculation uncovers the origin
of the quantum Hall plateau $\sigma_{xx}=0$ near the charge neutrality point by demonstrating
that a broad gap responsible for this quantum Hall state is formed because one of the valence-band
Landau levels rapidly goes up with increasing $B$. Very likely, the same mechanism describes
$\sigma_{xx}=0$ quantum Hall state observed previously in 2D semimetals with higher carrier
densities \cite{gusev}.\\

A financial support of this work by FAPESP, CNPq (Brazilian agencies),
RFBI and RAS programs "Fundamental researches in nanotechnology and
nanomaterials" and "Condensed matter quantum physics" is acknowledged.


\begin{thebibliography}{18}

\bibitem{novoselov}
%\bibitem{1}
K. S. Novoselov, A. K. Geim, S. V. Morozov, D. Jiang, M. I. Katsnelson,
I. V. Grigorieva, S. V. Dubonos, A. A. Firsov, Nature {\bf 438}, 197
(2005).

\bibitem{zhang}
%\bibitem{2}
Y. Zhang, Y.-W. Tan, H. L. Stormer, and P. Kim, Nature {\bf 438}, 201, (2005).

\bibitem{sarma}
%\bibitem{3}
S. Das Sarma, S. Adam, E. H. Hwang, and E. Rossi, Rev. Mod. Phys.,{\bf 83}, 407 (2011).

\bibitem{abanin}
%\bibitem{4}
D. A. Abanin, K. S. Novoselov, U. Zeitler, P. A. Lee, A. K. Geim,
and L. S. Levitov, Phys. Rev. Lett. {\bf 98}, 196806 (2007).

\bibitem{wiedmann}
%\bibitem{5}
S. Wiedmann, H. J. van Elferen, E. V. Kurganova, M. I. Katsnelson, A. J .M. Giesbers,
A. Veligura, B. J. van Wees, R. V. Gorbachev, K. S. Novoselov, J. C. Maan, and U. Zeitler,
Phys. Rev. B {\bf 84}, 115314 (2011).

\bibitem{dean}
%\bibitem{6}
C. R. Dean, A. F. Young, P. Cadden-Zimansky, L. Wang, H. Ren, K. Watanabe,
T. Taniguchi, P. Kim, J. Hone, and K. L. Shepard, Nature Phys. {\bf 7}, 693 (2011).

\bibitem{kvon}
%\bibitem{7}
Z. D. Kvon, E. B. Olshanetsky, D. A. Kozlov, et al., Pis'ma Zh.
Eksp. Teor. Fiz. {\bf 87}, 588 (2008) [JETP Lett. {\bf 87}, 502
(2008)].

\bibitem{dyakon}
M. I. Dyakonov and A. V. Khaetski, Sov. Phys. JETP {\bf 55}, 917 (1982).

\bibitem{brune}
C. Br\"une, C.X. Liu, E.G. Novik, E.M. Hankiewicz, H. Buhmann, Y.L. Chen, X.L. Qi,
Z.X. Shen, S.C. Zhang and L.W. Molenkamp, Phys. Rev. Lett. {\bf 106}, 126803 (2011).

\bibitem{kvon112}
%\bibitem{10}
Z. D. Kvon, E. B. Olshanetsky, E. G. Novik, D. A. Kozlov, N. N. Mikhailov,
I. O. Parm, and S. A. Dvoretsky, Phys. Rev. B {\bf 83}, 193304 (2011).

\bibitem{kittel}
C. Kittel, {\em Quantum Theory of Solids}, (Wiley, New York - London, 1963).

\bibitem{gusev}
%\bibitem{12}
G. M. Gusev, E. B. Olshanetsky, Z. D. Kvon, N. N. Mikhailov, S. A. Dvoretsky,
and J. C. Portal, Phys. Rev. Lett. {\bf 104}, 166401 (2010).

\bibitem{olshanetsky}
%\bibitem{13}
E.B. Olshanetsky, Z.D. Kvon, N.N. Mikhailov, E.G. Novik, I.O. Parm,
S.A. Dvoretsky, Sol.State Commun., to be published.

\bibitem{novik}
E. G. Novik, A. Pfeuffer-Jeschke, T. Jungwirth, V. Latussek,
C. R. Becker, G. Landwehr, H. Buhmann, and L. W. Molenkamp,
Phys. Rev. B {\bf 72}, 035321 (2005).

\bibitem{seeger}
K. Seeger, {\em Semiconductor physics, an introduction},
5th ed., (Springer, Berlin, 1997).

\bibitem{remark}
%\bibitem{16}
Of course, the numbering of the LLs can be chosen in a different way
within each subband and each spin sub-system. The numbering used in this paper is based
solely on the fact that $n$ is the number of the oscillatory function in Eq. (5).

\bibitem{konig}
M. K\"onig, H. Buhmann, L. W. Molenkamp, T. L. Hughes, C.-X. Liu, X.-L. Qi, and S.-C.
Zhang, J. Phys. Soc. Japan {\bf 77}, 031007 (2008).

\bibitem{konigthes}
M. K\"onig, Spin-related transport phenomena in HgTe-based quantum well structures.
Doctoral Thesis (W\"urzburg, 2007).


\end{thebibliography}
\end{document}